# New insights into the compressibility and high-pressure stability of Ni(CN)$_2$: A combined study of neutron diffraction, Raman spectroscopy, and inelastic neutron scattering

Sanjay K. Mishra[1], Ranjan Mittal[1,&], Mohamed Zbiri[2,*], Rekha Rao[1], Prabhatasree Goel[1], Simon J. Hibble[3,‡], Ann M. Chippindale[3,#], Thomas Hansen[2], Helmut Schober[2] and Samrath L. Chaplot[1]

[1]Solid State Physics Division, Bhabha Atomic Research Centre, Mumbai 400085, India
[2]Institut Laue-Langevin, 71 Avenue Des Martyrs, CS 20156, 38042 Grenoble Cedex 9, France
[3]Department of Chemistry, University of Reading, Whiteknights, Reading, Berks RG6 6AD, UK

Nickel cyanide is a layered material showing markedly anisotropic behaviour. High-pressure neutron diffraction measurements show that at pressures up to 20 kbar, compressibility is much higher in the direction perpendicular to the layers, $c$, than in the plane of the strongly chemically bonded metal-cyanide sheets. Detailed examination of the behaviour of the tetragonal lattice parameters, $a$ and $c$, as a function of pressure reveal regions in which large changes in slope occur, for example, in $c(P)$ at 1 kbar. The experimental pressure dependence of the volume data is fitted to a bulk modulus, $B_0$, of 1050 (20) kbar over the pressure range 0-1 kbar, and to 154 (2) kbar over the range 1-50 kbar. Raman spectroscopy measurements yield additional information on how the structure and bonding in the Ni(CN)$_2$ layers change with pressure and show that a phase change occurs at about 1 kbar. Changes in the Raman spectra are consistent with a phase change from Phase PI at ambient pressure, in which a square array of nickel atoms are linked by head-to-tail disordered cyanide groups in the Ni(CN)$_2$ sheets, to a new high-pressure phase, Phase PII, with ordered cyanide groups yielding sheets of $D_{4h}$ symmetry containing Ni(CN)$_4$ and Ni(NC)$_4$ groups. The Raman spectrum of phase PII closely resembles that of the related layered compound, Cu$_{1/2}$Ni$_{1/2}$(CN)$_2$, which has previously been shown to contain ordered C≡N groups. The phase change, PI to PII, is also observed in inelastic neutron scattering studies which show significant changes occurring in the phonon spectra as the pressure is raised from 0.3 to 1.5 kbar. These changes reflect the large reduction in the interlayer spacing which occurs as Phase PI transforms to Phase PII and the consequent increase in difficulty for out-of-plane atomic motions. It is notable that the collapse in the $c$ lattice parameter on forming the ordered PII form of Ni(CN)$_2$ brings its value, at 2.5 kbar, very close to that found in Cu$_{1/2}$Ni$_{1/2}$(CN)$_2$. Thus the ordering within the metal-cyanide layers appears to have a significant effect on interlayer forces. Raman spectra were measured up to 200 kbar before pressure release, when the sample reverted to the new phase, PII. Unlike other cyanide materials *e.g.* $Zn(CN)_2$ and $Ag_3Co(CN)_6$, which show an amorphization and/or a decomposition at much lower pressures (~100 kbar), $Ni(CN)_2$ can be recovered after pressurising to 200 kbar, albeit in a more ordered form. There is also evidence from the Raman spectra that a second phase transition occurs at ~70 kbar.

## I. INTRODUCTION

Metal cyanides, formed by linking together M(C≡N)$_x$ building blocks, can exhibit 1-, 2- and 3-dimensional structures. Disorder of carbon (C) and nitrogen (N) atoms within the C≡N groups is common, as both these atoms can occupy the same crystallographic positions due to their similar size and coordination preferences. For example, at ambient pressure, Zn(CN)$_2$, which is constructed from corner sharing Zn(CN)$_{2-x}$(NC)$_x$ tetrahedra, possesses a three-dimensional structure with cubic symmetry which can be described in either space group *Pn-3m* (for a model with C/N disorder)[1-3] or *P43m* (for an ordered model).[4] AgCN and AuCN consist of linear [–M(C≡N)M–] chains, again with head-to-tail disorder in the cyanide groups, packed into three-dimensional hexagonal structures.[5,6] Nickel cyanide, Ni(CN)$_2$, however, has a layered structure which is formed from linked square-planar Ni(CN)$_{4-x}$(NC)$_x$ units, with long-range order only in two dimensions (within the *a-b* plane) and no true periodicity along *c*.[7-9] Due to the 2-D nature of Ni(CN)$_2$, it has been proposed as an inorganic analogue of graphene, and is predicted to form nanotubular structures.[10] It is known to form intercalates and clathrates if molecular species are inserted between the layers.[11] The structure of one layer of Ni(CN)$_2$ is shown in Fig. 1(a), and the spatial relationship between adjacent layers in Fig. 1(b). The analysis of the powder X-ray diffraction data indicates that at ambient pressure the compound crystallizes with a tetragonal unit cell with *a* = 4.857 Å and *c* = 12.801 Å.[7] Ni(CN)$_2$ exhibits anomalous thermal expansion. The thermal expansion coefficient is found to be negative within the two-dimensionally connected sheets (in the *a-b* plane, $\alpha_a$ = –6.5×10$^{-6}$ K$^{-1}$) and has a high positive value perpendicular to the sheets ($\alpha_c$ = 61.8 × 10$^{-6}$ K$^{-1}$). The net result is an overall positive volume thermal expansion coefficient ($\alpha_V$ = 48.5 × 10$^{-6}$ K$^{-1}$).

In a previous study, we explored the origins of the thermal expansion behaviour of Ni(CN)$_2$ by measuring the temperature dependence of the phonon spectra using inelastic neutron scattering.[3] The results were analysed and interpreted using *ab initio* calculations. It was found that the experimental phonon spectra of Ni(CN)$_2$ do not show any significant temperature dependence, and the phonon modes of energy ∼2 meV are the principal contributors to the negative thermal expansion (NTE) observed within the Ni(CN)$_2$ layers. At 1 bar, the interactions between the layers appear to be weak but not entirely negligible since no soft modes are observed in the phonon spectra. These interlayer interactions might be expected to increase on application of pressure and the layered nature of the material makes it a suitable candidate for studying anisotropic compression with pressure. Thus high-pressure neutron diffraction experiments, along with inelastic neutron scattering and Raman measurements were carried out to understand the structural and dynamical behaviour of Ni(CN)$_2$, and gain new insights into its compressibility and stability under pressure.

## II. EXPERIMENTAL DETAILS

A polycrystalline sample of Ni(CN)$_2$ was prepared by heating Ni(CN)$_2$.3/2H$_2$O (sold by Alfa Aesar as Ni(CN)$_2$.4H$_2$O) under vacuum at 473 K for 48 h. Powder X-ray diffraction and IR spectroscopy showed that the hydrated nickel cyanide had been completely converted to anhydrous nickel cyanide and that Ni(CN)$_2$ was the only crystalline phase present. The resulting yellow powder was stored in a sealed glass ampoule prior to use, to prevent rehydration.

Neutron powder diffraction data for Ni(CN)$_2$ were measured at room temperature under pressures of up to 50 kbar using the high-flux D20 diffractometer[12] at the Institut Laue-Langevin (ILL), France. Low-pressure measurements (up to 2.7 kbar) were performed under hydrostatic conditions in a gas pressure cell, with argon gas as the pressure-transmitting medium, using neutrons of wavelength of 1.12 Å. For high-pressure measurements (up to 50 kbar), a Paris-Edinburgh (P-E) device[13] was employed using neutrons of wavelength of 2.41 Å. The sample, mixed with Pb metal as the pressure manometer, was loaded into an encapsulated Ti-Zr gasket filled with a 4:1 mixture of methanol-ethanol as the pressure medium, before pressing in the P-E device. The structural refinements were performed on the neutron diffraction data using the Rietveld refinement program FULLPROF.[14] In all the refinements, the background was defined by a sixth-order polynomial in 2$\theta$. A Thompson-Cox-Hastings pseudo-Voigt function with axial divergence asymmetry was chosen to define the profile shape for the neutron diffraction peaks.

Room temperature Raman spectroscopy measurements at high pressure were carried out on the polycrystalline sample of Ni(CN)$_2$ in a diamond anvil cell (Diacell B-05) with a 4:1 methanol-ethanol mixture as the pressure-transmitting medium. The pressure was measured using the ruby fluorescence technique. The sample was excited using a 532 nm excitation from a diode-pumped, frequency doubled solid-state laser with a power of ~15 mW. Scattered light was analysed using a home-made 0.9 m single monochromator, coupled with an edge filter and detected by a cooled CCD Raman data could not be collected between 0 and 1 kbar, due to the low accuracy of pressure determination in this pressure range.

Inelastic neutron scattering measurements were carried out at pressures up to 2.7 kbar on Ni(CN)$_2$ using the cold neutron, time-of-flight, time-focusing spectrometer, IN6, at the Institut Laue-Langevin (Grenoble, France) operating with an incident neutron wavelength $\lambda_i$ = 5.12 Å ($E_i$ = 3.12 meV), with a resolution of 80 μeV at the elastic line. About 7 grams of the polycrystalline sample of Ni(CN)$_2$ were compressed using argon in a gas pressure cell available at ILL. In order to measure the temperature dependence of $\frac{\partial \ln E_i}{\partial P}\left(=\frac{\Gamma_i}{B}\right)$ (where $\Gamma_i$, $E_i$, P and B are Grüneisen parameters, phonon energy, pressure and bulk modulus, respectively), the inelastic neutron scattering data of Ni(CN)$_2$ were collected at 200 and 300 K at a number of pressures up to 2.7 kbar. The measurements were first performed at 300 K, and the data were collected at ambient, 0.3, 1.9, and 2.7 kbar. After completing the measurements at 2.7 kbar, the pressure was released to ambient conditions. The temperature was then decreased to 200 K, and similar measurements were performed at the same pressures. The inelastic neutron scattering data were averaged over a scattering angle range of 10° to 113° and corrected for the contributions from argon at the respective pressures and for the background (including empty-cell measurements). A standard vanadium sample was used to calibrate the detectors. The data analysis was performed using ILL software tools, and the neutron cross-section weighted phonon densities of states were evaluated using standard ILL procedures.[15(a)] The relationship between the phonon density of states and the measured scattering function, $S(Q,E)$, is given in detail elsewhere.[3] The incoherent approximation[15(b)] was applied in the data analysis.

## III. RESULTS AND DISCUSSION

### A. High-Pressure Neutron Diffraction Study

Fig. 2(a) shows the neutron diffraction pattern of Ni(CN)$_2$ in a vanadium sample holder at ambient conditions. The diffraction pattern contains a mixture of sharp and broad peaks. The presence of well-defined sharp peaks reveals information about the periodic nature of the material; however, the broadening of certain reflections indicates stacking disorder. Similar diffraction patterns have been reported previously in other publications.[7,8]

Fig. 2(b) shows the pressure evolution of the diffraction patterns of Ni(CN)$_2$ in the gas pressure cell up to a pressure of 2.7 kbar. The diffraction patterns contain contributions from the sample and the aluminium (Al) container (marked with arrow). The anisotropic nature of the compressibility of Ni(CN)$_2$ is revealed by the separation of reflections which are coincident at ambient pressure. For example, above 1.2 kbar, the peaks with indices (300), (301), (206) and (008) at around $Q$ = 4

Å$^{-1}$, separate. The separation is due to the easier compressibility in the *c* direction as confirmed by the shift in the (103/004) reflection at $Q = 2$ Å$^{-1}$ to a higher *Q* value on application of pressure.

The maximum pressure achievable in the gas pressure cell is about 2.7 kbar. Further measurements at higher pressures were carried out using a Paris-Edinburgh (P-E) device. In addition, some of the measurements, below 2.7 kbar, were also repeated in the P-E device and the results are in good agreement with those from the gas pressure cell. Fig. 2(c) shows the evolution of the neutron diffraction patterns of Ni(CN)$_2$ in the P-E device. Peak shifts become even more obvious over this greater pressure range. In particular, above 0.5 kbar, the (004) reflection, at around $Q = 2$ Å$^{-1}$, moves towards higher *Q* values. We notice also that the intensity of the peaks decreases with increasing pressure.

Although the presence of a large number of stacking faults in Ni(CN)$_2$ precludes the determination of accurate atomic parameters, the variation of the lattice parameters with pressure could be determined from the diffraction data using Le Bail fitting. The quality of the fitting of the diffraction data is good (Fig. 3). This method of obtaining the lattice parameters was shown to be reproducible by carrying out the fitting using data sets in successive cycles of increasing and decreasing pressure.

Fig.4 depicts the evolution of the lattice parameters extracted from the refinement of the high-pressure diffraction data, collected using both the gas pressure cell and the P-E device. The value of the *a* lattice parameter, 4.835 Å, does not show an appreciable change up to 2.7 kbar. Further increase in pressure up to 50 kbar, which is the maximum pressure achieved, leads to a decrease in the *a* lattice parameter to 4.775 Å. The *c* lattice parameter also decreases as the pressure is raised to 50 kbar, and does so in a number of stages. The *c* lattice parameter first decreases slowly, up to just below 1 kbar and then falls more rapidly over the range 1 to 1.6 kbar (from 12.69 to 12.46 Å). There must be a change in the nature of the interactions between the layers because there is a sudden and substantial change in the compressibility at 1 kbar. Lattice parameter *c* continues to decrease up to 20 kbar (11.60 Å). The pressure dependence of the unit-cell volume shows a decrease over the entire pressure range of our study up to 50 kbar. It is notable that high-pressure Raman scattering measurements also change dramatically over the same range of pressure (see below).

Fig. 4 highlights the strongly anisotropic response under pressure of the lattice parameters with the variation of the interlayer spacing, *c*, reflecting a significantly larger axial contraction compared to the contraction in the *a-b* plane. In order to determine the bulk modulus at zero pressure, $B_0$, and its pressure derivative, B', the pressure-volume data were fitted by a third-order Birch-Murnaghan equation. A least-squares fit to the measured pressure dependence of the volume data in the range 0-1 kbar yields a bulk modulus, $B_0$, of 1050 (20) kbar, while over the remaining pressure range, 1-50 kbar, the $B_0$ value is 154 (2) kbar. The pressure derivative, B', was kept fixed at 4. The variation of the lattice parameters with pressure can be converted into compressibility values using the relationship: $K_l = -[(\ln l_P - \ln l_0)/(P - P_0)]$, where $l_P$ is the lattice parameter $l$ ($a$ or $c$) at pressure $P$, and $l_0$ is the lattice parameter $l$ at $P_0$. Over the pressure range 0 - 1 kbar, this leads to $K_a = +2.0 \times 10^{-4}$ kbar$^{-1}$ and $K_c = +3.5 \times 10^{-4}$ kbar$^{-1}$. Beyond the phase transition, over the pressure range 2 – 20 kbar, the values are $K_a = +3.5 \times 10^{-4}$ kbar$^{-1}$ and $K_c = +4.5 \times 10^{-3}$ kbar$^{-1}$. The magnitudes of $K_a$ and $K_c$ are comparable to those reported recently for layered silver(I) tricyanomethanide, although the latter material shows negative compressibility in two dimensions on application of pressure.[16]

For a proper understanding of the properties of the materials, the structural information, as obtained from diffraction techniques, should be supplemented and supported by other techniques. Inelastic neutron scattering offers a unique opportunity to reach a comprehensive conclusion on the dynamics. With this aim in mind, we performed Raman spectroscopy measurements and high-pressure inelastic neutron scattering experiments, which are discussed in the following sections.

**B. High-Pressure Raman Scattering Measurements**

Figs. 5 and 6 show the pressure evolution of the Raman spectra of Ni(CN)$_2$ up to ~200 kbar. At 1 bar, for Phase PI, the bands at ~2200 cm$^{-1}$ correspond to C≡N stretching modes and bands in the region 200-650 cm$^{-1}$ comprise Ni–CN/Ni–NC bending modes together with Ni–C/Ni–N stretching modes. On increasing the pressure from ambient to 1 kbar, dramatic changes are observed in all regions of the Raman spectra which indicate that a phase transition has occurred. These changes include a large decrease (of about 40 cm$^{-1}$) in the C≡N stretching-mode frequencies (red shift) (Fig.5 (b)), accompanied by an increase in separation of the two principal peaks and a reduction in their FWHM. In the low energy portion of the Raman spectrum, the relative intensities of the bands at ~275 cm$^{-1}$ and ~330 cm$^{-1}$ change dramatically, and the overall indication is that a phase transition, PI to PII, has occurred (see below). Above 1 kbar, the C≡N stretching modes undergo a blue shift up to 50 kbar (Fig. 5(b)). Around ~70 kbar, large changes occurring in the C≡N stretching modes are indicative of an additional phase transition (Figs 5(b) and 6(b)).

After releasing the pressure from 200 kbar, the Raman spectrum of the recovered sample was measured at ambient pressure. The spectrum resembles those observed at 1-3 kbar, indicating that Phase PII has formed and that it is stable at 1 bar. It is notable that, in contrast to other cyanide materials, *e.g.* $Zn(CN)_2$[17] and $Ag_3Co(CN)_6$[18, 19] which show an amorphization and/or a decomposition at much lower pressures (~100 kbar), $Ni(CN)_2$ can be recovered after pressurising to 200 kbar, albeit in a more ordered form. An interesting observation is that when the pressure is only raised as far as 4 kbar (Fig. 5(c)), Phase PII is formed, which then reconverts to Phase PI on the release of pressure back to 1 bar. Presumably in the initial transformation of PI to PII, defects remain and conversion is not wholly complete.

The PI to PII phase change on application of pressure can be ascribed to a change in C≡N order within the nickel-cyanide layers. In the monometallic transition-metal cyanides so far studied, definitive results, available for example in the cases of $Zn(CN)_2$,[1] $CuCN$,[20,21] $AgCN$[6] and $AuCN$,[5] reveal that in the normal forms prepared at atmospheric pressure there is head-to-tail cyanide disorder. Thus it is reasonable to suppose that head-to-tail cyanide disorder also occurs in $Ni(CN)_2$ in its low-pressure form, Phase PI, and this is supported by the Raman spectrum (*vide infra*). The strongest interactions, *i.e.* chemical bonds, occur within the nickel-cyanide layers and it is changes in the chemical bonding within the layers that have the greatest effect on the vibrational states that are sampled using Raman spectroscopy. Interactions between layers will slightly perturb these vibrational frequencies and might be expected to become more important as pressure increases. Thus the large changes which occur in the Raman spectrum at about 1 kbar, a relatively low pressure, are ascribed to changes in bonding within the nickel-cyanide layers. The sharpening of the ν(C≡N) bands at ~2200 cm$^{-1}$ is ascribed to the formation of layers in which the cyanide groups are ordered. Possible ordering schemes within $Ni(CN)_2$ layers, which may be applicable to Phase PII, have been discussed by Mo and Kaxiras,[10] who, in their work on metal-cyanide nanotubes, determined the energies of individual nickel-cyanide sheets for three different ordered arrangements (Fig. 7). DFT calculations found that the sheet of $C_{2v}$ symmetry assembled from *cis*-$Ni(CN)_2(NC)_2$ units lies lowest in energy. The $D_{4h}$ sheet built from $Ni(CN)_4$ and $Ni(NC)_4$ units has a slightly higher calculated energy (0.07 eV per $Ni(CN)_2$ unit) than the *cis*-$Ni(CN)_2(NC)_2$ modification whilst the $D_{2h}$ sheet containing *trans*-$Ni(CN)_2(NC)_2$ units has the highest calculated energy (0.17 eV per $Ni(CN)_2$ unit above the *cis* form). As we show below, it is the $D_{4h}$ sheet structure which forms at about 1kbar (Phase PII). [It should be noted that the calculated relative energies[10] are for individual sheets, whereas in the solid, the interactions between sheets may be sufficient to change the energy order of the three different forms. There is clearly a significant change in the interactions between the layers on going from phases PI to PII as seen in the large change in the $c$ lattice parameter (Fig. 4)].

Although there are no known examples of M(CN)$_2$ sheets containing M(CN)$_2$(NC)$_2$ units, there is a recently reported example containing M(CN)$_4$ and M(NC)$_4$ units.[22] The copper-nickel cyanide, Cu$_{1/2}$Ni$_{1/2}$(CN)$_2$, is isostructural with Ni(CN)$_2$ with both metal centres having square-planar coordination, but has been shown unequivocally to contain ordered Ni(CN)$_4$ and Cu(NC)$_4$ groups within the M(CN)$_2$ layers to form sheets with $D_{4h}$ symmetry[22] (Fig. 1(a)). Comparison of the Raman spectrum of Cu$_{1/2}$Ni$_{1/2}$(CN)$_2$ with those found for the PI and PII forms of Ni(CN)$_2$ (Fig.8) is highly suggestive that pressure has induced a phase transition in Ni(CN)$_2$ in which a disordered layer transforms to one with $D_{4h}$ symmetry. The clear splitting of the two ν(C≡N) Raman active stretches and the low-energy region in the Raman spectrum of Cu$_{1/2}$Ni$_{1/2}$(CN)$_2$ bear very close resemblance to the features seen for Ni(CN)$_2$ in its PII form. The spectra of PII and Cu$_{1/2}$Ni$_{1/2}$(CN)$_2$ also show an absence of modes in the region 550 − 650 cm$^{-1}$, which are seen for Ni(CN)$_2$ in its PI form.

Group theoretical analysis of the three sheets shown in Fig. 7 predicts for the $C_{2v}$ structure: 12 Raman active bands ($3A_1 + 3A_2 + 2B_1 + 4B_2$); for the $D_{2h}$ structure: 12 Raman active bands ($4A_g + 4B_{1g} + 2B_{2g} + 2B_{3g}$); and for the $D_{4h}$ structure: 8 Raman active bands ($2A_{1g} + 2B_{1g} + 2B_{2g} + 2E_g$). For an isolated Ni(CN)$_4^{2-}$ unit with $D_{4h}$ symmetry,[23] 7 Raman active bands are predicted ($2A_{1g} + 2B_{1g} + 2B_{2g} + E_g$). Using the relative positions and intensities from the Raman spectrum of this isolated unit, six bands for Cu$_{1/2}$Ni$_{1/2}$(CN)$_2$ and hence Ni(CN)$_2$-PII can be assigned as shown in Fig. 8.[‡] The remaining two bands, $B_{2g}$ (predicted for the isolated unit to be at 109 cm$^{-1}$)[23] and the additional $E_g$ (predicted when the square-planar units are connected into a $D_{4h}$ sheet) might be expected to lie below 200 cm$^{-1}$ and hence are not observed. It is difficult to reconcile the observed Raman spectrum with the group theoretical predictions for the $C_{2v}$ and $D_{2h}$ structures.

Pressure-induced ordering of C≡N groups has been observed previously. For example, Zn(CN)$_2$ undergoes a phase transition at about 10 kbar from a cubic to an orthorhombic structure,[17, 24] and it has been suggested that this orthorhombic phase is likely to show CN ordering, unlike the ambient-pressure phase.[1] In addition, over the pressure range 0-1200 kbar in the Prussian Blue analogue, K$_{0.4}$Fe$_4$[Cr(CN)$_6$]$_{2.8}$.16H$_2$O, in which Cr$^{III}$–CN–Fe$^{II}$ linkages isomerise to the Cr$^{III}$–NC–Fe$^{II}$ form.[25] This isomerization is reversible on releasing the pressure when low pressures are applied, but after pressing at ~12 kbar, a metastable phase is formed which does not re-isomerise to the original form.

## C. High-Pressure Inelastic Neutron Scattering Measurements

The phonon spectra of Ni(CN)$_2$, measured at ambient pressure, 0.3, 1.5 and 2.7 kbar, at 200 K and 300 K are shown in Fig. 9. The large amount of scattering from the high-pressure cell only allowed measurement of the phonon spectra up to 40 meV. The spectra consist of broad peaks centred on 7, 9, 17, 21 and 30 meV. The intensity of the broad peaks decreases and the peaks below 20 meV shift significantly towards higher energies with increasing pressure; the others do not change by a significant amount. These observations are consistent with our previous assignment of these low-energy modes to motions involving out-of-plane atomic motions.[3] These motions become more constrained under pressure as the layers are squeezed together.

Our previous *ab-initio* calculations[3] for Ni(CN)$_2$ show that the contributions from Ni atoms extend up to 75 meV, while C and N atoms contribute in the whole phonon spectra range up to 280 meV. The Ni atoms provide the main contribution to the low-energy phonon spectra up to 40 meV. It is found that all the modes shift towards higher energies as pressure increases. This is what would be expected for compounds exhibiting positive thermal expansion behaviour. The experimental data (Fig. 9) allow extraction of the mode Grüneisen parameters $\left( \frac{\Gamma}{B} = \frac{1}{E} \frac{\partial E}{\partial P} \right)$ (where B, E and P are the bulk modulus, phonon energy and pressure, respectively). As shown in Fig. 9, the Grüneisen parameters, as extracted for Ni(CN)$_2$ from the pressure dependence of the phonon spectra (0 to 2.7 kbar) at 200 K and 300 K, are found to have significantly different values. The difference seems to be significant in the low-energy region below 10 meV. We have previously reported[26] high-pressure inelastic neutron scattering measurements of Zn(CN)$_2$ up to 2.8 kbar at 165 K and 225 K. In contrast to Ni(CN)$_2$, the Grüneisen parameters for Zn(CN)$_2$, as extracted from the high-pressure data up to 2.8 kbar, were found to be very similar at 165 and 225 K[26] (Fig. 10) illustrating that although the two cyanide materials both contain M−C≡N−M units, the way that these are further connected has a profound effect on the lattice dynamics and their pressure and temperature dependence.

The volume thermal expansion coefficient, $\alpha_V$, for Ni(CN)$_2$ is 48.5 × 10$^{-6}$ K$^{-1}$, while Zn(CN)$_2$ has an isotropic NTE coefficient, $\alpha_V$, of −51 × 10$^{-6}$ K$^{-1}$. Although the signs of the volume thermal expansion coefficients for the two compounds are different, the values have similar magnitude, $|\alpha_V| \sim 50 \times 10^{-6}$ K$^{-1}$ and hence the average values of $|\Gamma(E)/B|$ are expected to be nearly the same in both compounds.

However, on comparing $\Gamma(E)/B$ for both compounds (Fig. 10), it can be seen that $|\Gamma(E)/B|$ for Ni(CN)$_2$ has twice the magnitude of $|\Gamma(E)/B|$ for Zn(CN)$_2$.[17] We have also extracted $\Gamma(E)/B$, at 200 K and 300 K, from the two different pressure ranges of our measurements (Fig. 11). The values of $\Gamma(E)/B$ obtained from the two different pressure regimes are found to differ significantly.

These findings from the phonon spectra are very interesting in that they provide further evidence of a pressure-induced phase transition in Ni(CN)$_2$. Associated with the change in CN order within the layers, and a consequent softening of the C≡N stretching modes by ~40 cm$^{-1}$ at about 1 kbar, there appears to be a change in interlayer interaction and a sudden relaxation in the *c* direction. Previous high-pressure Raman measurements on Ag$_3$Co(CN)$_6$ also showed that the C≡N stretching modes soften by about 50 cm$^{-1}$ as pressure is increased to 2 kbar.[16] High-pressure neutron diffraction studies[19] on Ag$_3$Co(CN)$_6$ also indicate a phase transition at the same pressure. Our diffraction data (Fig. 4) also show a change in the slope of the lattice parameters, as well as in the volume, at about 1 kbar. This lattice-parameter behaviour is yet further evidence of a phase transition at moderate pressure values of about 1 kbar.

## IV. Conclusions

In summary, we report the high-pressure evolution in the structure and dynamics of Ni(CN)$_2$ using a combination of neutron diffraction, inelastic neutron scattering and Raman spectroscopy.

That the powder diffraction pattern of Ni(CN)$_2$ does not change drastically at moderate pressures and that the interlayer spacing of Ni(CN)$_2$ above 1kbar is similar in value to that measured for the recently characterised compound, Cu$_{1/2}$Ni$_{1/2}$(CN)$_2$ (at 1 bar),[22] provides compelling evidence for retention of the simple layered structure under these conditions. A change in geometry around some or all of the nickel from square planar to tetrahedral on application of pressure would be accompanied by significant changes in the diffraction pattern, rather than the gradual evolution of the pattern and the lattice parameters as observed here. Indeed polymorphs of ZnNi(CN)$_4$ which contain tetrahedral Zn(NC)$_4$ square-planar Ni(CN)$_4$ units and consequently have 3-D framework structures have very different diffraction patterns.[27,28]

Overall, the response of the lattice parameters of Ni(CN)$_2$ to applied pressure is strongly anisotropic with a significantly larger contraction occurring along the *c* axis than in the *a-b* plane. This behaviour is as a result of the existence of strong chemical bonding within the Ni(CN)$_2$ layers, parallel to the *a-b* plane (the relatively incompressible part of the structure), and no chemical bonding between the layers. Above 1 kbar, the diffraction measurements show a change in the slope of the variation of the $c$ lattice parameter with pressure, which is the signature of a phase transition. From high-pressure Raman measurements and, by comparison with, $Cu_{1/2}Ni_{1/2}(CN)_2$, we ascribe this phase transition (Phase PI to Phase PII) to a change in CN ordering within the nickel-cyanide layers with retention of square-planar coordination around nickel. A second phase transition at ~70 kbar is evident in the cyanide-stretching region of the Raman spectrum but no structural conclusions can be drawn.

The pressure dependence of the phonon spectra of $Ni(CN)_2$, from the inelastic neutron scattering measurements, also highlights the occurrence of the pressure-induced phase transition, PI to PII, at about 1 kbar and supports the conclusion from our previous work[3] that motions perpendicular to the layers are major contributors to the density of states at low energies.

‡ [Footnote. A very weak peak seen at ~400 cm$^{-1}$ in PI and $Cu_{1/2}Ni_{1/2}(CN)_2$ probably arises because of the loss of the centre of symmetry when the layers are stacked together. It is notable that a very intense band with $\underline{E}_u$ symmetry is predicted at ~400 cm$^{-1}$ in the $Ni(CN)_4^{2-}$ unit[23]].

**ACKNOWLEDGEMENTS**

The Institut Laue-Langevin (ILL) facility, Grenoble, France, is acknowledged for providing beam time on the IN6 spectrometer and on the D20 diffractometer. The authors acknowledge the use of the Raman scattering facility at BARC, Mumbai, India for the high-pressure Raman measurements, and the use of powder X-ray diffraction instrumentation in the Chemical Analysis Facility (CAF) of the University of Reading, UK, for preliminary characterization of the nickel-cyanide samples.

&rmittal@barc.gov.in
*zbiri@ill.fr
#a.m.chippindale@rdg.ac.uk
‡s.j.hibble@jesus.ox.ac.uk

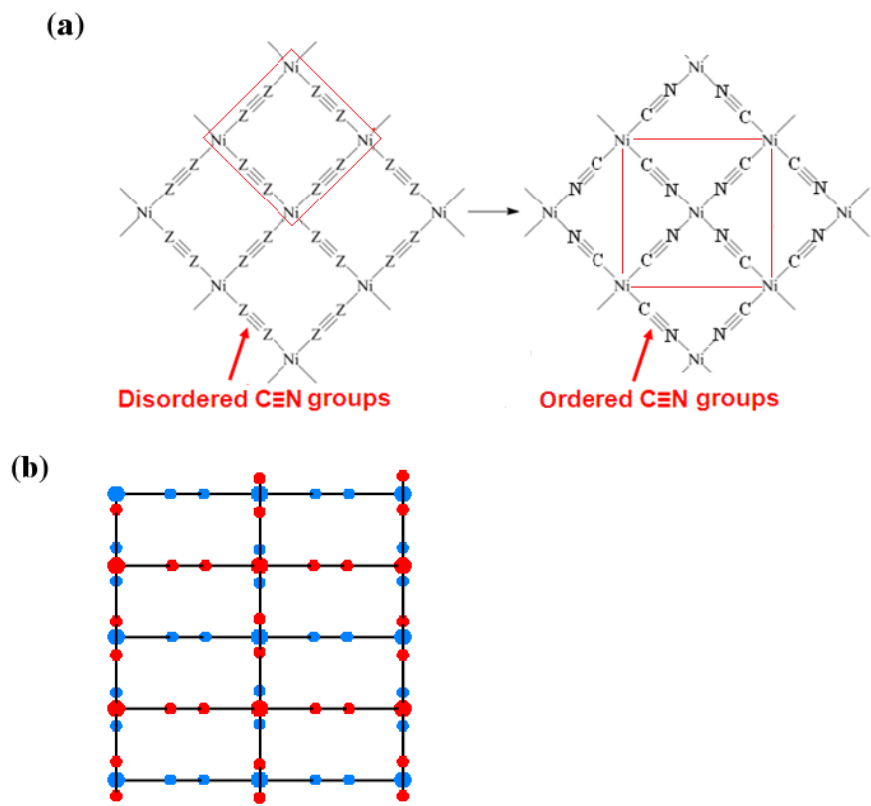

**Fig. 1**(a) The structure of one layer of Ni(CN)$_2$ in the *a-b* plane. In the left panel, the C/N atoms, represented by Z, are head-to-tail disordered, whilst in the right panel, the cyanide groups are shown as ordered. (b) View along the *c*-axis showing the fixed stacking relationship between adjacent Ni(CN)$_2$ layers (shown in blue and red). Cyanide groups (small spheres) always lie above the nickel atoms (large spheres) in the layer below.

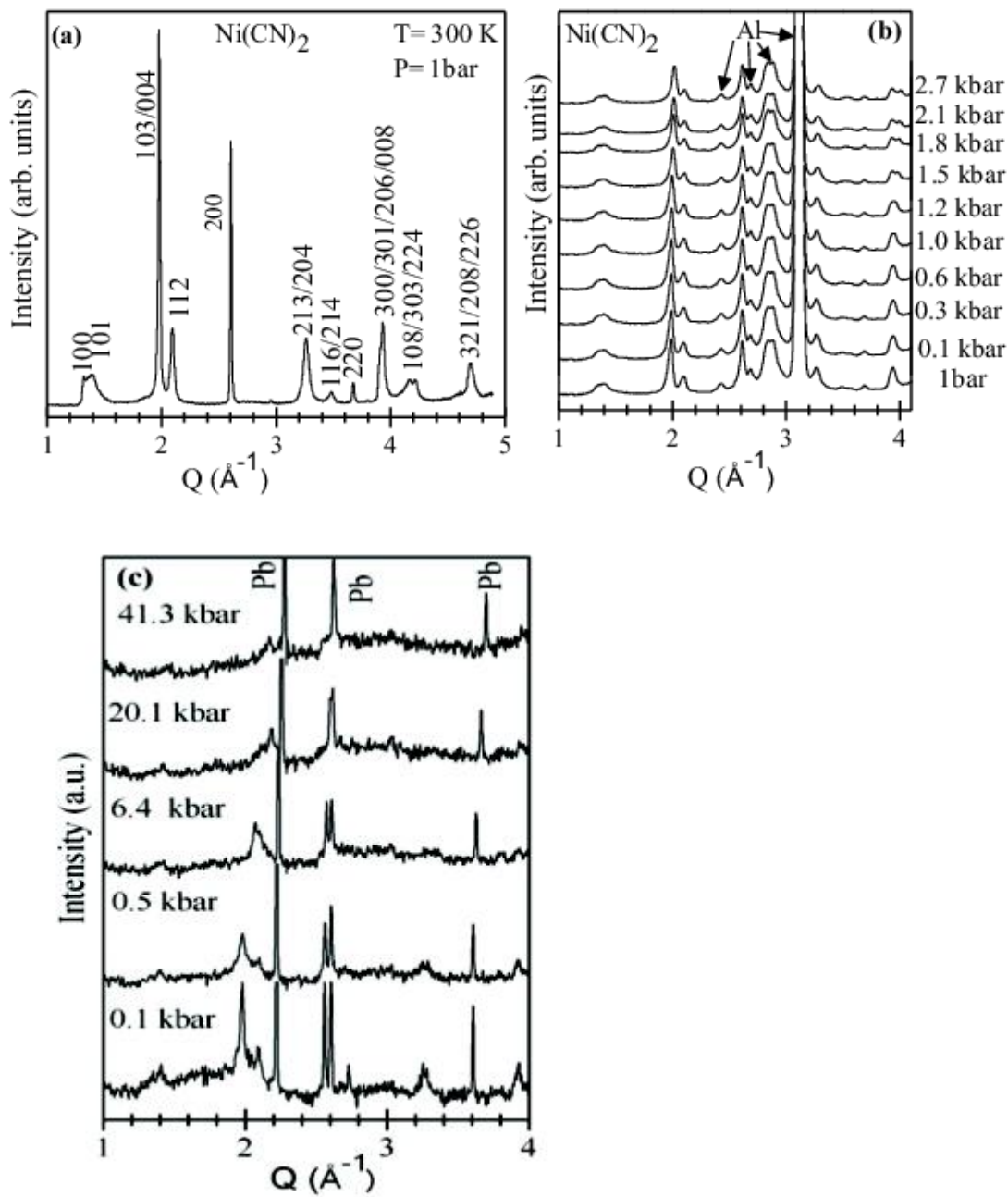

**Fig. 2.** (a) Powder neutron diffraction pattern of Ni(CN)$_2$ under ambient conditions. (b) Evolution of neutron diffraction patterns of Ni(CN)$_2$ at selected pressures up to 2.7 kbar, measured using the argon gas pressure cell. The peaks labelled Al originate from the aluminium sample holder. (c) Evolution of neutron diffraction patterns of Ni(CN)$_2$ at selected pressure points measured in a Paris-Edinburgh device.

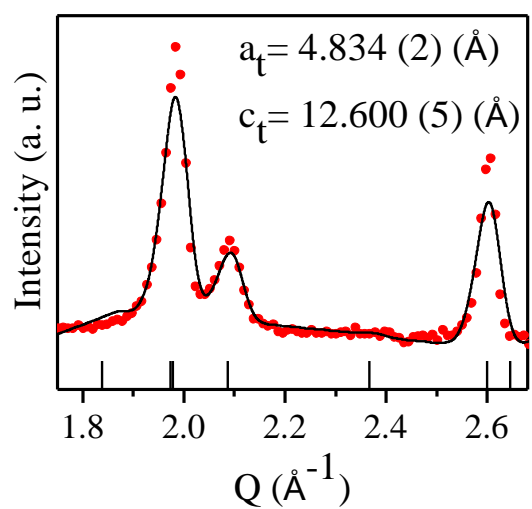

**Fig. 3.** Fitting of neutron diffraction data at 1.2 kbar using the Le Bail approach.

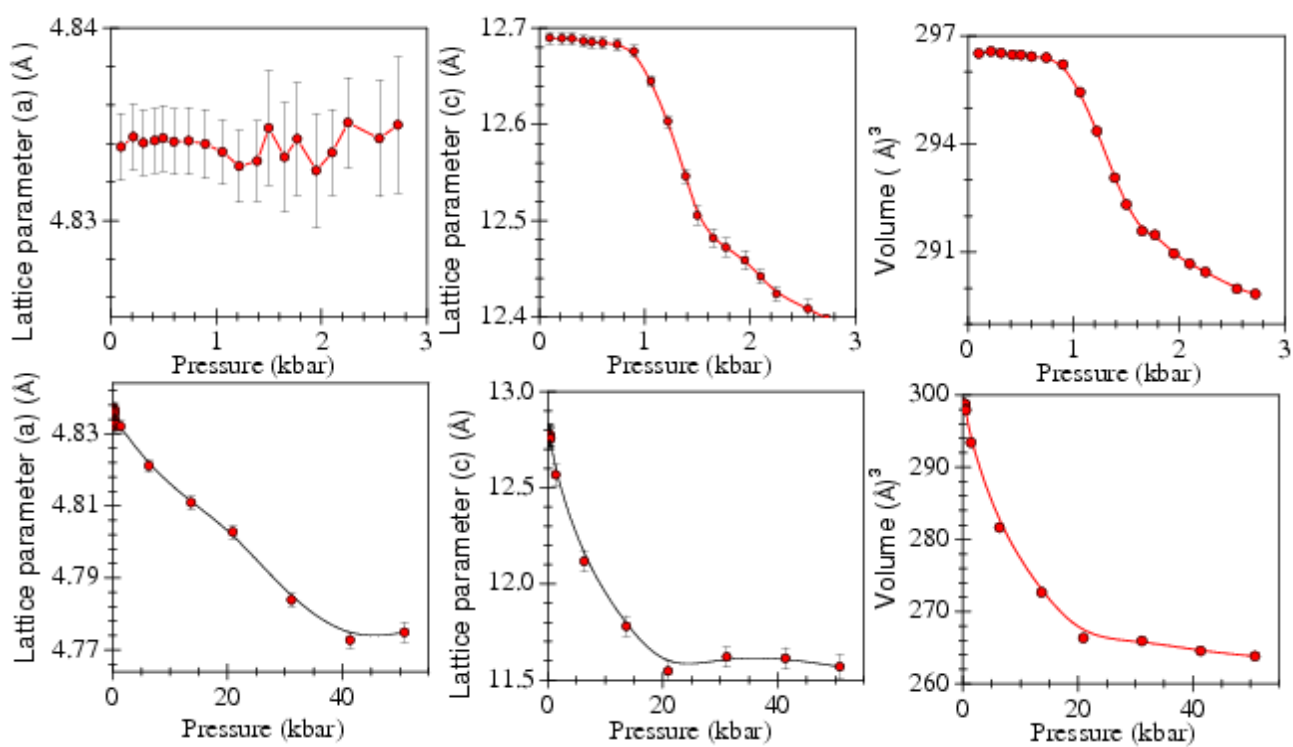

**Fig. 4.** The pressure dependence of the structural parameters, *a*, *c* and *V,* of Ni(CN)$_2$ measured up to 2.7 kbar using the argon gas pressure cell (top panel), and up to 50 kbar using the Paris-Edinburgh (P-E) device (bottom panel).

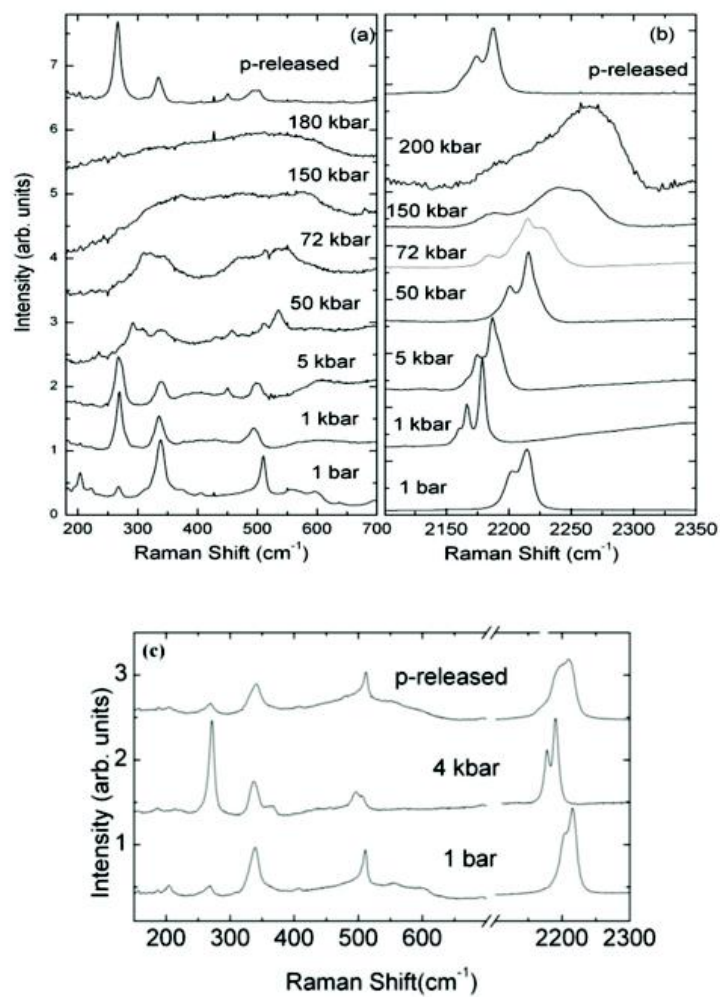

**Fig. 5.** Room-temperature Raman spectra of Ni(CN)$_2$ at various pressures in the region of (a) the Ni–CN/Ni–NC bending and Ni–C/Ni–N stretching and (b) the C≡N stretching modes (Note: The Raman spectra labelled 'p-released' are measured at ambient pressure after releasing the pressure from 200 kbar). (c) The Raman spectra of a fresh sample of Ni(CN)$_2$ after compression up to 4 kbar, followed by decompression to ambient pressure.

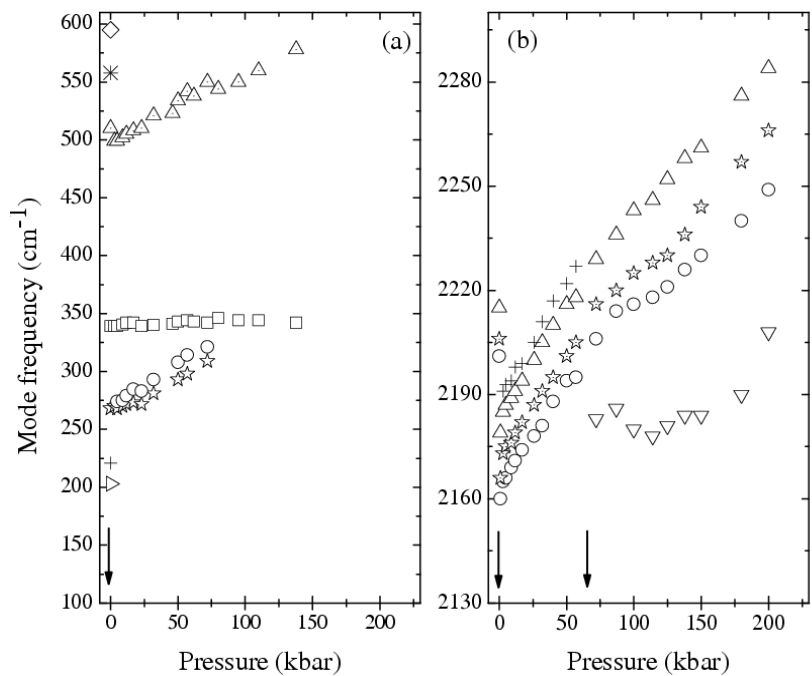

**Fig. 6.** The pressure dependence of the Raman shifts of (a) Ni–CN/Ni–NC bending and Ni–C/Ni–N stretching modes and (b) C≡N stretching modes. The arrows indicate the transition pressures at ~1 and ~70 kbar.

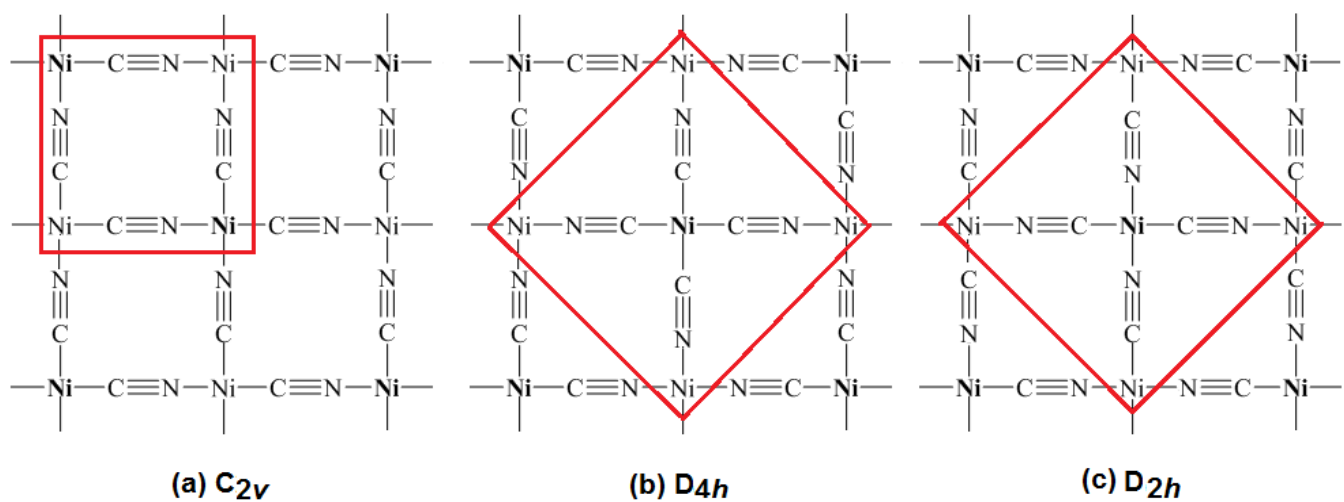

**Fig. 7.** Three ordered variants of Ni(CN)$_2$ sheets with (a) $C_{2v}$, (b) $D_{4h}$ and (c) $D_{2h}$ symmetry. The repeating units within the sheets are outlined in red.

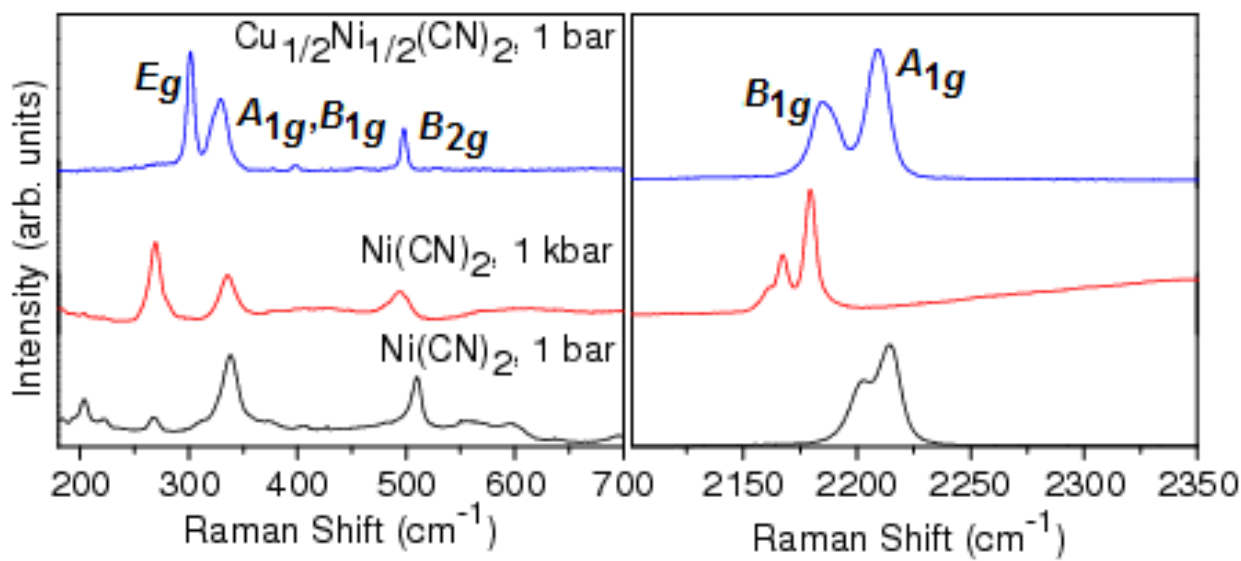

**Fig. 8.** Room temperature Raman spectra of Ni(CN)$_2$ (Phase PI) (black line) and Cu$_{1/2}$Ni$_{1/2}$(CN)$_2$[22] (blue line) at ambient pressure. Note the similarity of the spectrum of Cu$_{1/2}$Ni$_{1/2}$(CN)$_2$ to that of Ni(CN)$_2$ at 1 kbar (Phase PII) (red line).

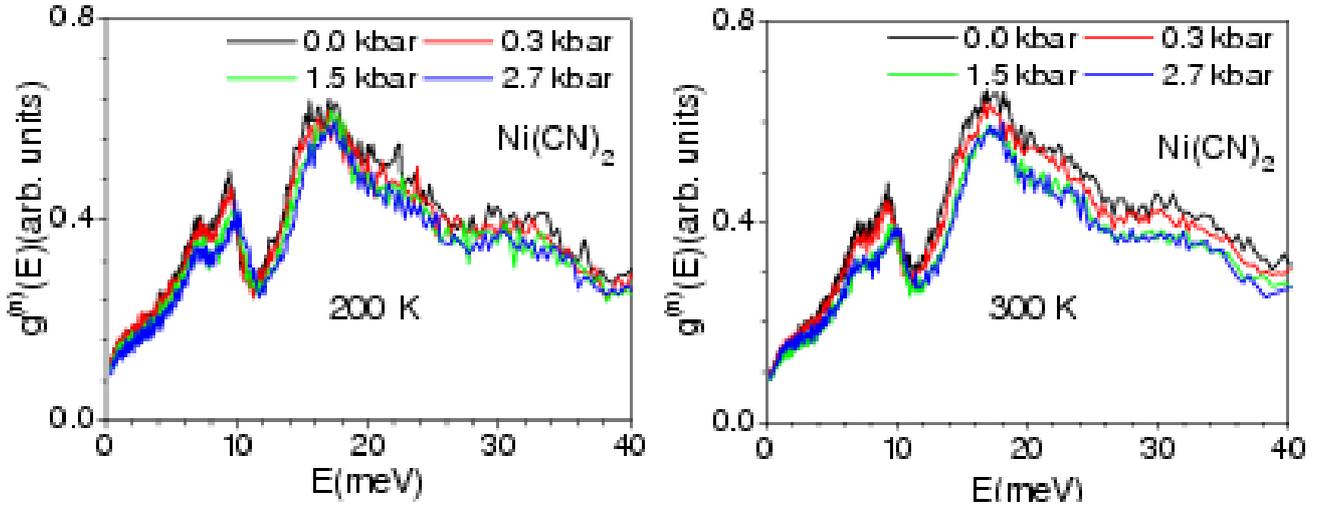

**Fig.9.** Pressure dependence of the phonon spectra of Ni(CN)$_2$ from inelastic neutron scattering measurements using the IN6 spectrometer at 200 K (left) and 300 K (right).

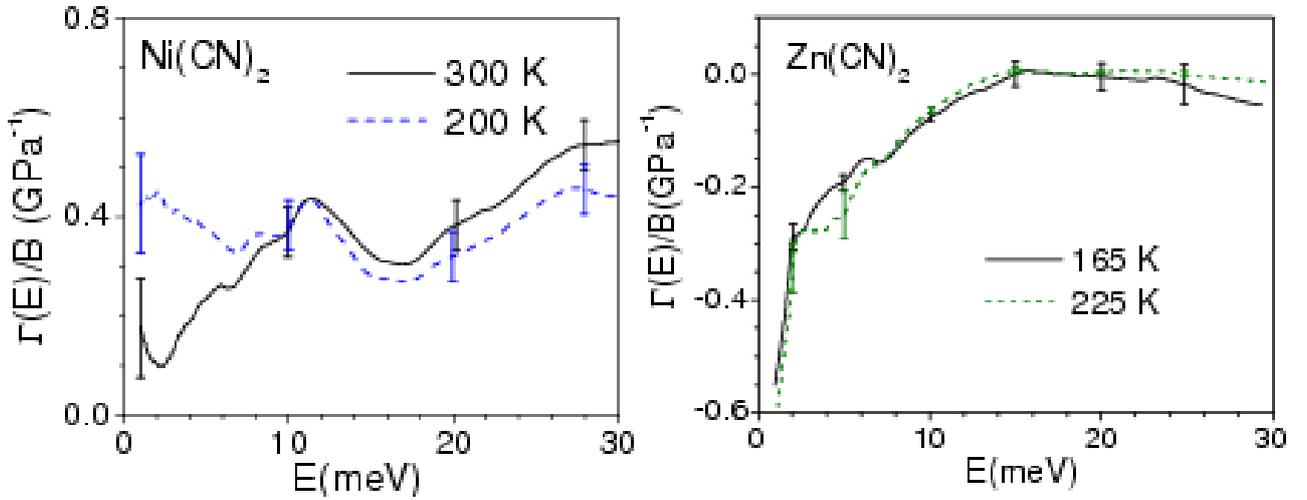

**Fig. 10.** The experimental $\frac{\Gamma_i}{B}$ as a function of phonon energy $E$ (averaged over the whole Brillouin zone) for Ni(CN)$_2$ and Zn(CN)$_2$.[24] The $\frac{\Gamma_i}{B}$ values at 200 K and 300 K were determined using the phonon density of states at $P = 0$ and 2.7 kbar, which represent the averages over the whole range in this study.

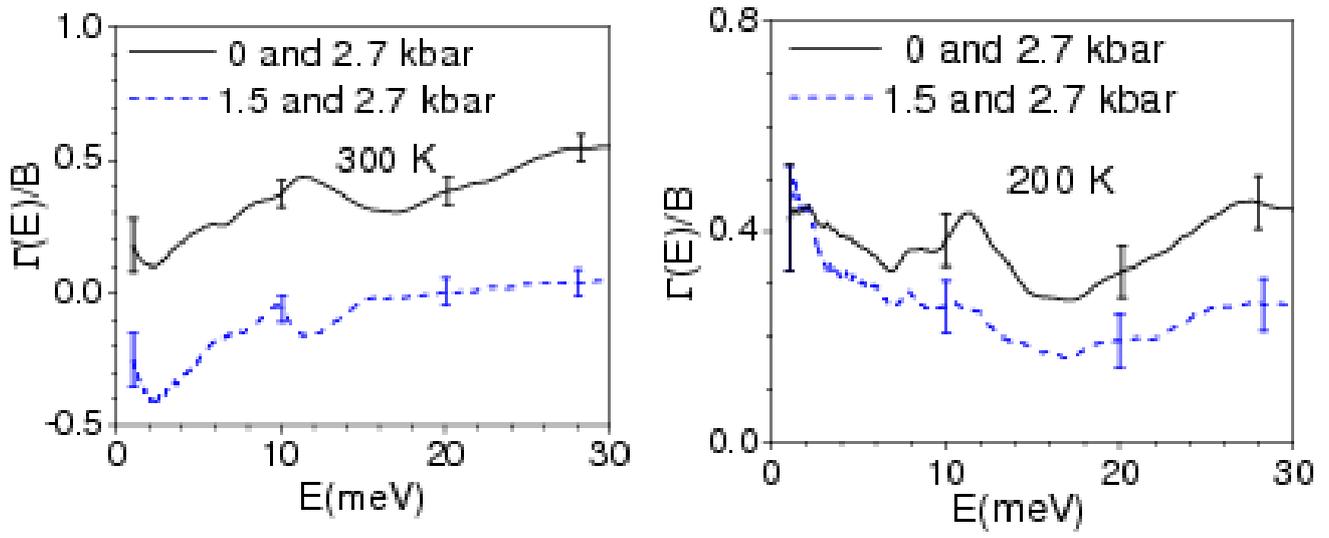

**Fig. 11.** The experimental $\frac{\Gamma}{B}$ as a function of phonon energy $E$ (averaged over the whole Brillouin zone) for Ni(CN)$_2$. The full and dashed lines represent the pressure shifts obtained from the measurements carried out in the pressure range of 0 to 2.7 kbar, and 1.5 to 2.7 kbar, respectively.